\documentclass[twocolumn,showpacs,preprintnumbers,amsmath,amssymb]{revtex4-1} 
\usepackage{graphicx}% Include figure files
\usepackage{dcolumn}% Align table columns on decimal point
\usepackage{bm}% bold math
\usepackage{cases}
\usepackage[tight,hang]{subfigure}

\def\diadic#1{\overline{\overline{#1}}} 
\def\ni{\noindent}
\def\sig{\sigma}

\def\br{\bf{r}}

\def\bg{\bf{g}}

\def\bR{\bf{R}}

\def\heps{{\hat \epsilon}}

\def\heps{{\diadic{\epsilon}}}

\def\beq{\begin{equation}} 
\def\eeq{\end{equation}}

\begin{document} 
\draft 
%\large 
 
%\twocolumn[\hsize\textwidth\columnwidth\hsize\csname 

\title{Theory of strains in auxetic materials}
\author{Raphael Blumenfeld$^{1,2}$, Sam F. Edwards$^2$}

%\address{1. Earth Science and Engineering and Inst. of Shock Physics, Imperial College London, London SW7 2AZ, UK \\
%2. Biological and Soft Systems, Cavendish Laboratory, J J Thomson Avenue, Cambridge CB3 0HE, UK}

\affiliation{1. ISP and ESE, Imperial College, London SW7 2AZ, UK \\
2. Cavendish Laboratory, JJ Thomson Avenue, Cambridge CB3 0HE, UK}

%\maketitle

\date{\today}

\begin{abstract} 

This paper is dedicated to Prof.   Jacques Friedel, an inspirational scientist and a great man. His excellence and clear vision led to significant advances in theoretical physics, which spilled into material science and technological applications. His fundamental theoretical work on commonplace materials has become classic. We can think of no better tribute to Friedel than to apply a fundamental analysis in his spirit to a peculiar class of materials - auxetic materials.
Auxetic materials, or negative-PoissonÕs-ratio materials, are important technologically and fascinating theoretically. When loaded by external stresses, their internal strains are governed by correlated motion of internal structural degrees of freedom. The modelling of such materials is mainly based on ordered structures, despite existence  of auxetic behaviour in disordered structures and the advantage in manufacturing disordered structures for most applications.
We describe here a first-principles expression for strains in disordered such materials, based on insight from a family of `iso-auxetic' structures. These are structures, consisting of internal structural elements, which we name `auxetons', whose inter-element forces can be computed from statics alone. Iso-auxetic structures make it possible not only to identify the mechanisms that give rise to auxeticity, but also to write down the explicit dependence of the strain rate on the local structure, which is valid to all auxetic materials. 
It is argued that stresses give rise to strains via two mechanisms: auxeton rotations and auxeton expansion / contraction.  The former depends on the stress via a local fabric tensor, which we define explicitly for 2D systems. The latter depends on the stress via an expansion tensor. Whether a material exhibits auxetic behaviour or not depends on the interplay between these two fields.
This description has two major advantages: it applies to any auxeton-based system, however disordered, and it goes beyond conventional elasticity theory, providing an explicit expression for general auxetic strains and outlining the relevant equations.

\end{abstract}
%\pacs{64.60.Ak, 05.10.c 61.90.+d}
%\narrowtext  

\maketitle

\bigskip
\ni {\bf I. Introduction by Sam F. Edwards}
\bigskip

At the end of the war, which had isolated France from the English speaking world, several French scientists moved to UK universities, in particular to study solid state theory. The outstanding person in the UK at the time was Nevill Mott in Bristol and Jacques Friedel moved to Bristol to work in Mott's group. 
At the time, the field of theoretical physics was moving into the use of field theory to elucidate elementary particle theory, a direction favoured in Cambridge University and in London. The Bristol group, however, specialised in electronic studies, an area that Friedel preferred. 
I remember his papers at that time, which had a wonderful clarity and discussed down-to-earth type of problems. It was refreshingly in stark contrast to the renormalisation theory, which was the fashion in quantum field theory at the time. 

Sometime later, Mott moved to Cambridge and Jacques returned to Paris. This reminds me my first conference in Paris, where I gave my first paper. It was nonsense, I regret to say, for it tried to separate green functions for the real and imaginary parts of the wave function. Fortunately, none of the attendants in that conference exposed it. 

Anyway, I recall Friedel giving wonderful lectures in cambridge, where his work was held at very high esteem. Years later Cambridge University awarded him an Honorary Doctor of Science and I had the pleasant task of arranging a dinner for him.  Friedel was also involved in setting up the European physical society, where I was active, and I recall him giving valuable advice on its structure.

A central sociological problem in theoretical physics is to choose the problem to work, for there are many brilliant people working at the forefront of the field. Thinking of Friedel's work on electronics in parts of systems, it occurred to me that one should be able to do statistical mechanics on continuous systems in contrast to particulate systems. With my coauthor here, Raphael Blumenfeld, I have developed this idea by studying the entropy of particulate systems in the continuum. For example, in conventional thermal systems the entropy $S$ is a function of pressure, volume, energy and number of particles, $S(E,P,V,N)$ and one of the most useful concepts it leads to is the temperature $T=\partial E/\partial S$. We have applied these to granular systems where the entropy is due to configurational disorder and the volume takes the role of the energy. Consequently, the analogue of temperature is the `Compactivity' $X_0=\partial V/\partial S$. There are other quantities that dictate the states of granular matter, the simplest being the response of stresses to the entropy, $X = \partial \sig/\partial S$, which we called the Angoricity (note that the Angoricity is in fact a tensor). An even richer and more general 'thermodynamics' is required when we study mixtures.

This paper is dedicated to Jacques Friedel and, in the spirit of the close relations of his theoretical works with real materials, we can think of no better tribute to him than to present a fundamental theory that aims  to understand the physics of a peculiar class of materials - auxetic materials.

`
\bigskip
\ni {\bf II. General introduction}
\bigskip

Auxetic materials, i.e. materials with negative Poisson's ratio, expand when stretched and contract when compressed, in contrast to most conventional materials. This is due to correlated degrees of freedom in the internal elements that theses materials are made of. These elements are reversibly foldable and, in effect, can be regarded as the basic constituents of cellular solids. In the following, we call these foldable elements `auxetons'. 
Macroscopic auxetic structures can be manufactured of polymers\cite{La87} or metals\cite{La87,FrLa88}. 
They can exist on a range of length-scales and, in particular, can be constructed out of molecular building blocks\cite{WeEd99,WuWe04a,WuWe04b}. 
Auxetic materials are useful in applications requiring high shear to bulk moduli or compactification on impact, e.g. for energy absorbing materials and bullet-proof armours.

Both natural\cite{Hoho89,NuSi69,CaEv89,EvCa89} and man-made\cite{Al85,Ev89} auxetic materials have been discovered, made and studied. 
Much of the theoretical analysis, however, is carried out on ordered models, such as two-dimensional inverted cell honeycombs. Although models of the auxeticity phenomenon in ordered structures is convenient for analysis purposes, the ubiquity of disordered such materials and the little existing understanding of deformations in the presence of disorder require a more general theory.
Here we describe such a theory, based on a recent suggestion made in \cite{Bl05}. 

The aims of this paper are the following. 
First, we describe a new family of disordered auxetic structures, called iso-auxetic (IA) structures, for which it is possible to identify clearly the basic strain mechanisms. 
Second, we show that elasticity theory is not necessary for the description of auxeticity, implying that using negative Poisson's ratio as a descriptor has a limited utility. 
Third, we present an explicit expression for the auxetic strain in terms of local expansive and rotational fields. In this expression, the fields are coupled to the stress through well-defined tensors, which we discuss. 
Fourth, we show that auxeton rotations are essential to the understanding of the global behaviour and that the rotational field can be modelled without resorting to non-symmetric stresses. This obviates models based on Cosserat theory\cite{Cosserat}. 

The paper is structured as follows. 
We first introduce the new family of IA structures. These are structures whose inter-auxeton forces can be determined from statics alone. This property distinguishes IA from more conventional auxetic structures, which we term  elasto-auxetic (EA). Specifically, the stress field equations of isostaticity theory differ significantly from those of conventional elasticity in that they are based on local stress-structure relations, as opposed to the usual stress-strain relations\cite{Wietal96,Wietal97,Caetal98}. 
We next describe an extension of a recent result for yield of granular systems to IA structures and write down explicitly the IA strain equation in terms of two local fields: an expansive and a rotational.
It is then argued that the mechanism for auxeticity depends only on these two fields and is therefore independent of the particular way that the structure transmits stresses, whether isostatically or elastically. Hence, the auxetic expression for the strain is valid for {\it all} auxetic materials. This, in turn, implies that general auxeticity needs to be described by a theory that goes beyond elasticity. 
A particular implication of this conclusion is that negative Poisson's ratio in auxetic materials should be regarded only as a descriptor of the ratio of strains in perpendicular directions, not as a ratio of elastic moduli. 
We also argue that, although the way the form of the strain expression is the same for all auxetic materials, the strains developing in IA structures differ markedly from those developing in EA structures under the same loading conditions. 
We conclude with a discussion of the results.

\bigskip
{\bf II. Iso-auxetic structures}
\bigskip

In the following discussion, we consider planar auxetic materials made of 2D elementary units that connect to their neighbours at exactly three  points. We call these elements `auxetons'. Aiming at a theory of disordered materials, we do not require that the auxetons be identical, nor that the system possess any type of symmetry, translational or otherwise. Rather, we consider systems whose auxetons comprise a mixture irregular sizes, shapes and orientations. A wide variety of such structures can be constructed, some of which are illustrated in figure 1.  

\begin{figure}["here"]
\begin{center}
\includegraphics[width=6cm]{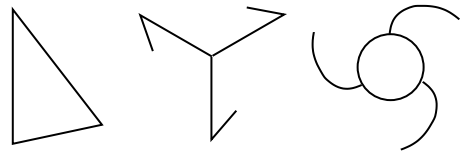}
\caption{\it Examples of auxetons made of three-contact building blocks. Each auxeton can expand and rotate when forces are applied to its ends, termed `contacts' in the text.}
\label{fig1}  
\end{center}
\end{figure}

We constrain our auxetons to have three `contacts' with their neighbours and connect these contacts by imaginary straight lines into triangles (the dashed blue lines in figure 2). This construction results in a planar graph of triangles, connected at their vertices. The triangles enclose polygons, which we call in the following cells. 

\begin{figure}["here"]
\begin{center}
\includegraphics[width=6cm]{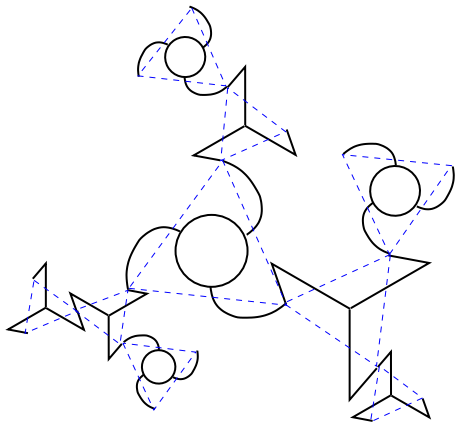}
\caption{\it A section of a disordered auxetic structure, made of joining auxetons at their contacts. The contacts are joined by straight lines (blue dashed) into a triangle. These triangles are then used to characterise the contact network in a well-defined manner.}
\label{fig2}  
\end{center}
\end{figure}

When loaded by external forces, the auxetons transmit those to one another through `inter-auxeton' forces. The contacts between auxetons may or may not be free-jointed. One expects the latter to be more common, in which case a contact can  support a certain threshold of torque moment without yielding. This gives rise to a finite overall stress threshold for straining the material. 
Consider then a structure, made of $N(\gg 1)$ auxetons, stressed {\it below} the yield threshold by a set of external forces. Below the yield threshold, the system is in mechanical equilibrium and all the inter-auxeton forces and torques are balanced. Since every auxeton has three contacts then the number of contacts is $3N/2 + O(\sqrt{N})$, where the latter term is a boundary correction, which can be neglected for $N\gg 1$. Since each contact transfers one force vectors, there are overall $3N$ force components. These can be determined uniquely by the three balance equations for every auxeton - one of torque and two of force components. 
It follows that this structure is statically determinate, or {\it isostatic}. Hence the name iso-auxetic.
A familiar textbook statically determinate system is that of a ladder on a frictional floor leaning against a frictionless wall. The forces that the wall and the floor apply to the ladder can be determined uniquely from its  three balance equations. 
It is important to note that, as in the case of the ladder problem, the determination of the discrete inter-auxeton forces requires no knowledge whatever of the elastic properties of neither the auxetons nor the contacts. Since the stress field is nothing but a continuos representation of the large number inter-auxeton forces, it must reflect the nature of the discrete solution and therefore be independent of local elastic moduli. It follows that elasticity theory, which does rely on knowledge of the elastic moduli, is inapplicable for IA structures.

For later discussion, it is useful to recall the continuum 2D stress equations of isostaticity theory - the theory of stresses in isostatic structures,

\begin{eqnarray}
\frac{\partial\sig_{ij}}{\partial x_i} = g_j \label{ForceBal} \\
\sig_{ij} = \sig_{ji} \label{TorqueBal}  \\
Q_{ij}\sig_{ij} = 0 \label{ConstEq} 
\label{StressEqs}
\end{eqnarray}
Eqs. (\ref{ForceBal}) and (\ref{TorqueBal}) represent force and torque balances, respectively, with $\sig$ the stress tensor and $\bg$ external forces. Eq. (\ref{ConstEq}) is a constitutive relation between the static stress and the local structure, which is characterised by a symmetric fabric tensor $Q$\cite{BaBl02,Bl04a,BaBl05}. This replaces the  stress-strain relations in conventional elasticity and is indeed independent of the elastic moduli of the material.

In most known isostatic systems these equations are hyperbolic, leading to solutions that `propagate' along characteristic paths in the material. This means that the response to a localised force source in 2D is generically a pair of force chains. In contrast, EA materials respond to localised sources by `dispersing' the stress field in all directions, subject to local stress-strain relations. The differnce between the two types of solutions stems from the nature of the stress field equations - while the equations of elasticity theory are {\it elliptic}, eqs. (ForceBal)-(ConstEq) of isostaticity theory are {\it hyperbolic}.
The different stress transmission is bound to affect macroscopic behaviour, as will be discussed below.

The global auxetic behaviour is the result of local folding and unfolding of auxetons when stressed. This local response is independent of whether the rest of the structure is isostatic or not, it is only dependent on the local expansion and rotation of the auxetons. This leads to the conclusion that the strain can be written in terms of local expansion and rotation fields regardless of the isostatic or elastic nature of the material. This conclusion is  significant for two reasons. One is that, in IA, the stress is independent of elastic moduli. The other is, that in IA we can write the strain explicitly in terms of the expansion and rotation of auxetons, which means that the same expression holds for EA materials. This gives insight into the description of auxeticity in general and in particular into the coupling between the local strain and the local structure. Additionally, this suggests that elastic constants need not play as major a role as in conventional materials. Another important implication is that the negative Poisson's ratio, which such materials exhibit, is only a descriptor of the ratio of perpendicular strains and is of little use in terms of describing bulk elastic moduli because these cannot be obtained simple homogenisation of small-scale regions. 
It is also worthwhile to note, before we continue, that this description should apply not only to all auxetic materials made of foldable auxetons, but also to those made of rigid ones\cite{Gretal05}. 

Before we proceed, we must comment on a much debated issue: whether  or not auxeticity theory necessitates resorting to Cosserat stress theory\cite{Cosserat}, which allows for existence of a non-symmetric stress tensor. 
This is not a question of formalism, but rather of the underlying physics. A symmetric stress tensor means that residual torque moments vanish on the continuum length-scales. Differently put, it means that there are no external couple moments on the system that require balancing mechanically by the mechanical stress field. By letting the stress tensor be non-symmetric on macroscopic scales, Cosserat theory implies that there exist external couples that the stress must balance. 
Thus, a theory that invokes only symmetric stresses does not resort to such additional input and must be preferable for modelling of large-scale auxetic behaviour. For this reason we prefer the above formulation, which includes (eq. (\ref{TorqueBal}).

Nevertheless, it is important to point out that a symmetric stress tensor can still allow existence of local rotational fields of the material upon straining. In other words, although the stress field must be symmetric under no external couples, the strain field may have non-symmetric contributions.
Indeed, local such contribution arise from rotation of auxetons and it is at the core of auxetic behaviour, as will be discussed in the next section.

\bigskip
{\bf III. Auxetic strain and field equations}
\bigskip

To relate the strain to the local structure one has to have first a quantitative description of the structure, however disordered. Such a descriptor is the fabric tensor $Q_{ij}$ of eq. (\ref{ConstEq}). This tensor plays a key role in modelling auxetic strains, as will be seen below. 
Consider a disordered structure of auxetons, comprising an arbitrary mixture of elements, such as those shown in figure 1. The model to be described below has been discussed initially in \cite{Bl05} and it is general in that it applies to any arbitrary structure of the above auxetons. Specifically, the disorder can involve both auxetons of different sizes and of different shapes. 
Connecting the three contact points around each auxeton by straight lines, as described above, the plane is tiled into a network of triangles of different sizes and shapes, all interconnecting at their vertices - the contact points. The triangles enclose polygons, which we will call in the following cells. 
According to Euler relation\cite{Euler}, a system of $N\gg 1$ such auxetons would have two auxeton per cell. This value has small corrections from boundary auxetons (due to unshared contacts), but this correction is of order $\sim 1/\sqrt{N}$ and therefore negligible.
All triangle edge are assigned directions, making them into vectors $\br$ that circulate the triangles in the clockwise direction (figure 3).

\begin{figure}["here"]
\begin{center}
\includegraphics[width=6cm]{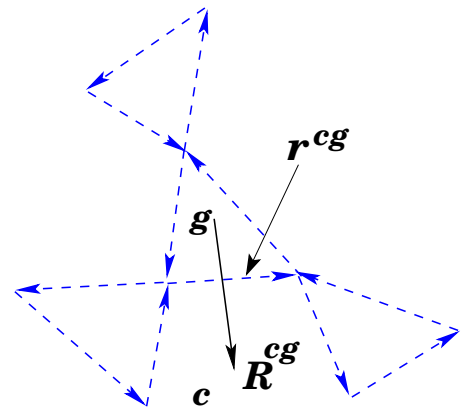}
\caption{\it Characterisation of the auxeton structure in 2D. We make the edges of the representative triangle $g$ into vectors, $\br^{cg}$, by assigning the edges a direction such that they circulate around the triangle in the anti-clockwise direction. The vector $\bR^{cg}$ extends from the centroid of the grain contacts to the centroid of an adjacent  cell $c$. }
\label{fig3}  
\end{center}
\end{figure}

\ni Every triangle is assigned a centroid, defined as the  mean position vector of its three vertices. Similarly, every cell is assigned a centroid, defined as the  mean position vector of the contacts (triangle vertices) that surround it. In mechanical equilibrium, the cell polygons must be convex to be stable. This means that, a vector $\bR^{cg}$ extending from the centroid of triangle $g$ to the centroid of one of its neighbour cells $c$, intersects one of the triangle edge vectors, which we can index $\br^{cg}$ (figure 3). 

The vectors $\bR^{cg}$ and $\br^{cg}$ can be regarded as the diagonals of a quadrilateral, called `quadron', which plays a significant role in granular and cellular physics\cite{BlEd03,BlEd06,Blaus07}. Each quadron is associated uniquely with a pair $cg$ (see figure 4). 
This construction allows us to quantify the local structure by quantifying the shape of every $cg$-quadron tensorially as the outer product 

\begin{figure}["here"]
\begin{center}
\includegraphics[width=6cm]{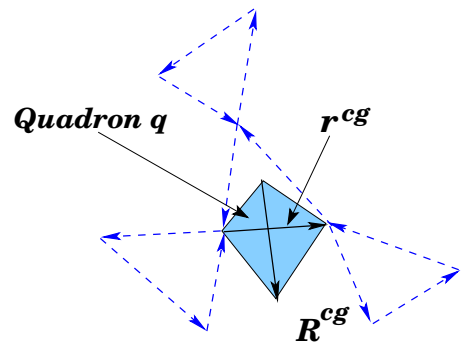}
\caption{{\it Quadron tessellation in 2D}. The vectors $\br^{cg}$ and $\bR^{cg}$ make the diagonals of the $cg$-quadron. The quadrons are the elementary units that tessellate the system.  The quadron shape is quantified by a local structure tensor, $C^{cg}_{ij}= r^{cg}_i R^{cg}_j$.}
\label{fig4}  
\end{center}
\end{figure}

\beq
C^{cg}_{ij} = r^{cg}_i R^{cg}_j
\label{StrucTensQ}
\eeq
The tensor appearing in the isostaticity stress equations is the symmetric part of $C^{cg}$, summed over the cells around triangle $g$

\beq
Q^{g} = \frac{1}{2}\heps^{-1}\cdot\left( \sum_{c\ around\ g} C^{cg} + \left(C^{cg}\right)^T \right)\cdot\heps
\label{FabTensQ}
\eeq
where $\heps$ is the $\pi/2$ rotation matrix in the plane (the Levi-Civita matrix) and $C^T$ is the transpose of $C$. 

Armed with a quantitative description of the local structure, it is possible now to relate it to the strain. 
Suppose that the structure is in mechanical equilibrium under a set of external forces and these forces are increased. Eventually, the system crosses what is known as the yield surface and it starts deforming. As will become clearer below, whether the deformation is auxetic or not depends on the structure of the auxetons, their configuration and the magnitude of the local stresses. The aim of the following is to describe the equations that govern the strain, given the local structure and the local stress. 

\ni Central to the model is the observation that only auxeton rotation and expansion (or shrinking, which can be regarded as negative expansion) can give rise to displacement. The expansion corresponds to pure folding / unfolding of auxetons. 
Thus, the local strain $e$, due to changes in shape and volumes of the triangles, can be written as a superposition of a rigid triangle rotation, $e^{rot}$, and triangle (non-uniform) expansion, $e^{ex}$. 
For example, auxetic materials composed of rigid auxetons, such as those studied in \cite{Gretal05}, can be described by $e^{rot}$ alone. 
In the following we consider only the symmetrised strain, but there is no reason why the treatment should not apply to non-symmetric strains equally well.
Note that the dependence of the strain on the stress is only through the responses of both these modes of motion to local stresses. This is, in fact, the main difference between this theory and elasticity-based descriptions that relate directly the strain to the stress. 
 
\ni The (symmetrised) strain due to rotation at the centroid of auxeton $g$ is given directly by the tensor $Q^g$\cite{BaBl03},

\beq
e^{rot,g}_{ij} = Q^g_{ijkl}\theta^g_{kl}
\label{RotStrn}
\eeq
where $\theta^g_{kl}$ is its angle of rotation, which depends on the local stress. Eq. (\ref{RotStrn}) is written so that it holds both in 2D and in 3D. In 3D, this expression is symmetric under exchange of the indices $i$ and $j$, but anti-symmetric under exchange of $k$ and $l$. This is due to the anti-symmetric nature of the description of the axes of rotations $kl$. 
In 2D, there is only one axis of rotation, perpendicular to the plane, and the indices $kl$ are redundant, which reduces $Q_{ijkl}$ to the tensor $Q_{ij}$ of eq. (\ref{FabTensQ}).
This expression has been derived first in \cite{BaBl03} for granular media, where it gives rise  to dilation. It comprises the {\it only} relevant contribution to the strain when the auxetons are rigid and, as such, should also describe well the systems discussed in \cite{Gretal05}. For what follows, it is important to note the observation in \cite{BaBl02,Bl04a} that $Q$ is a measure of the local rotational (or chiral) deviation of the auxeton from a global zero average. This rotation is best quantified by the sign of Tr$\{ Q\}$.

When elements can also fold and unfold, their expansions depend on the local stress. Significantly, there is no reason to expect that auxeton expansions be isotropic; depending on the choice of shape and the local structure around them, auxetons may expand differently in different directions. 
The expansive strain rate can be related directly to the local stress via

\beq
e^{exp,g}_{ij} = E^g_{ijkl}\sig^g_{kl} 
\label{ExpStrn}
\eeq
where the non-isotropic expansion can be modelled into $E^g$ and different auxeton shapes would be described by different such matrices.  
Limiting the description to symmetric strains imposes some constraints on the local expansion tensor $E$, making its properties similar to those of the conventional compliance matrix in linear elasticity. However, such similarity would not exist for non-symmetric strains. For example, for such strains, $E$ need not be symmetric under exchange of $i$ and $j$. 
It is important to note that, whilst the strain may have non-symmetric components, for example to describe large-scale vorticity, the stress cannot if there are no external couples to balance the excessive torque. This is one of the reasons that the following theory cannot be mapped readily to elasticity theory, nor to Cosserat theory. 

The total strain can be written then as 

\beq
e^{g}_{ij} = E^g_{ijkl}\sig^g_{kl} + Q^g_{ijkl}\theta_{kl}\left(\sig^g\right)
\label{TotStrnG}
\eeq
This relation is reminiscent of the yield equations in granular systems\cite{BaBl03}, but for two important differences. One is that, in granular systems, the rotating elements (the grains) can also slide relative to neighbours, a mechanism that auxetons do not possess, which gives rise to an additional, plasticity-like term. The other difference is that auxetons can fold and unfold (the $E$-dependent term), which rigid grains cannot. 

Relation (\ref{TotStrnG}) makes good sense on the auxeton level. However, to be of use to materials that contain many auxetons, it must be coarse-grained (homogenised) to the continuum. To this end, one must average it over small volumes, containing sufficiently  many auxetons. The expansive term on the right hand side of (\ref{TotStrnG}) gives no problems - one can average $E$ and $\sig$ independently to obtain a continuum-scale contribution. This is no different than the practice in conventional elasticity and plasticity models. 

In contrast, the rotational term requires a careful consideration. Coarse-graining over the rotation field of a region of volume $V$, $\langle\theta\rangle = (1/V)\sum_g\theta^g$, can be carried out by replacing the volume average by a surface sum (or integral, for large enough regions), using Stokes theorem. This leads immediately to the observation that the contribution to such an average comes only from the boundary of the region. Hence, if the system does not rotate globally, then the rotation per auxeton decays fast as the averaging volume increases and the macroscopic rotation has a zero average. 
It turns out that the tensor $Q$ possesses exactly the same property. Since this tensor measure the local chiral fluctuation of an element, an average over a region decays to the global zero average at exactly the same rate as $\langle\theta\rangle$. 

On the face of it, these two observations may seem to imply that the rotational contribution to the strain vanishes on large scales. This, however, is not the case. The reason is that both $\theta$ and $Q$ possess the same local anti-correlations: when one auxeton rotates in one direction, elements in contact with it are more likely to rotate in the opposite, rather than in the same, direction. Similarly, if the tensor $Q^g$ measures the rotation of an auxeton at a given direction, nearest-neighbours of $g$ are more likely than not to have $Q$'s whose trace has the opposite sign. This anti-correlation means that, while each of these terms averages to zero independently over increasing regions, their product $\langle Q_{ijkl}\theta_{kl}\rangle$ adds {\it constructively} over nearest neighbours, leading to a finite large-scale average. 
It is exactly this average that leads to measurable bulk strain due to rotations of rigid particles in granular systems (dilation). 
We therefore conclude that eq. (\ref{TotStrnG}) has a well defined homogenised large-scale version

\beq
e_{ij} = E_{ijkl}\sig_{kl} + Q_{ijkl}\theta_{kl}\left(\sig\right)
\label{TotStrn}
\eeq
The only remaining question is how to derive local continuous expressions for the rotational term. This can be done, using the renormalisation approach taken in \cite{Bl04b}. 
A word of caution: the existence of a macro-scale continuous theory does not imply that the strain is auxetic. Relation (\ref{TotStrn}) gives the {\it correct} dependence of the strain on the local fields, but whether the ratio between perpendicular strains (Poisson's ratio) is negative or positive depends on the relative contribution of the two terms on the right hand side of this relation. 

The advantage of relation (\ref{TotStrn}) is that it identifies the precise role that the local structure plays in the coupling to the strain and the stress. As such, it is an important ingredient in the field equations of auxeticity theory. To complete the theory for IA structures, one still needs the local rotational response  to the local stress, $\theta_{kl}(\sig)$. This relation is still missing and work to derive it is ongoing. 
Thus, the full set of auxeticity field equations in $d$ dimensions consist of: \\
(i) $d(d+1)/2$ balance eqs. (\ref{ForceBal})-(\ref{ConstEq}); \\
(ii) $d(d-1)/2$ strain eqs. (\ref{TotStrn}); \\ 
(iii) $d(d-1)/2$ rotation - stress response relations, $\theta_{kl}(\sig)$. \\ 
As in any theory, constitutive information is required.  For the theory described here, this comprises the constitutive tensors $E$ and $Q$, which could be obtained either phenomenologically or modelled theoretically for specific structures. 

The solution for quasi-static deformation then proceeds as follows. First, one solves for the stress field from eqs. (\ref{ForceBal})-(\ref{ConstEq}). From this solution one finds the local rotational field $\theta_{kl}(\sig)$, using the local rotation - stress relation. Substitution of the rotational field, the constitutive fabric tensor $Q$ and the expansion tensor $E$ into eq. (\ref{TotStrn}) one then derives the total local strain. 

Since the dependence of the strain on local rotation and expansion of elements is valid regardless of the stress state, then all this theory, but for the stress field equations, applies to {\it any} auxetic material. In particular, it applies to  EA materials, where the stress equations should then be replaced by those of elasticity theory. In other words only the closure relation (\ref{ConstEq}) is replaced by Saint tenant compatibility conditions\cite{Venant}, supplemented with phenomenological or modelled expression for the stress-strain relations.
 
\bigskip
{\bf IV. Discussion and conclusions}
\bigskip

To conclude, we have described a theory for strains in auxetic materials. 
The theory's main contribution is the explicit relation between the local auxetic strain and the local rotation and expansion of auxetons - the elementary building blocks of auxetic materials. This is a refinement of the currently existing elasticity theory which lumps these two contributions together into a stress-strain relation. The identification of these strain mechanisms makes it possible to eventually arrive such a relation, since the local magnitudes of auxeton rotations and expansions do depend on the local stress. However, the explicit decomposition to rotation and expansion give insight into the correct symmetryies and details of such a stress-strain relation.

Furthermore, using elasticity theory for IA would lead to erroneous results, which originate from two sources. Firstly, the stress state cannot be derived from elasticity theory and is likely to exhibit non-uniform force-chain-like fields. Secondly, the rotational and expansion responses to the stress are of completely different nature. For example, the averaging properties of $Q$ and $E$ are completely different - while the latter has a well-defined macroscopic homogenised value, the latter does not. This is despite both terms having homogenised large-scale contributions.

A significant implication of the above is that all auxetic, whether IA or EA, must follow the universal strain eq. (\ref{TotStrn}). However, the stress state, which determines the local rotation and expansion of auxetons, depends on the correct stress description and this may vary between different families of materials - isostaticity theory for IA and elasticity theory for EA. This then leads to the intriguing conclusion that the auxetic behaviour of IA and EA materials should be markedly different, with the former exhibiting more non-uniform local auxetic behaviour. 

It is emphasised that eq. (\ref{TotStrn}) does not ensure auxeticity, but rather it describes correctly the strain as a function of the local rotational and expansive fields. Whether the material exhibits a bulk negative Poisson's ratio depends on the different contributions of the two terms in the strain relation.


\begin{thebibliography}{99}

\bibitem{La87} R.S. Lakes, 
{\it Foam structures with a negative PoissonÕs ratio}, Science {\bf 235}, 1038 (1987).
\bibitem{FrLa88} E.A. Friis, R.S. Lakes, J.B. Park, 
{\it Negative PoissonÕs ratio polymeric and metallic materials}, J. Mater. Sci., {\it 23}, 4406 (1988).
\bibitem{WeEd99} G. Y. Wei and S. F. Edwards, 
{\it Effective elastic properties of composites of ellipsoids (I). Nearly spherical inclusions}, PHYSICA {\bf A 264}, 388 (1999).
\bibitem{WuWe04a} H. M. Wu and G. Y. Wei, 
{\it Molecular design of new kinds of auxetic polymers and networks},  Chinese J. Pol. Sci. {\bf 22}, 355 (2004).
\bibitem{WuWe04b} H. M. Wu and G. Y. Wei, 
{\it Molecular design of several types of self-assembly auxetic networks},  Acta Pol. Sin. {\bf 2}, 201 (2004). 
\bibitem{Hoho89} F. Homand-Etienne, R. Houpert, 
{\it Thermally induced microcracking in granites: characterization and analysis}, Int. J. Rock Mech. Min. Sci and Geomech., Abstract {\it 26}, 125 (1989).
\bibitem{NuSi69} A. Nur, G. Simmons, 
{\it The effect of saturation on velocity in low porosity rocks. Earth Planet}, Sci. Lett., {\bf 7}, 183 (1969).
\bibitem{CaEv89} B.D. Caddock, K.E. Evans, {\it Microporous materials with negative PoissonÕs ratio: I. Microstructure and mechanical properties}, J. Phys. D., Appl. Phys. {\bf 22}, 1877 (1989).
\bibitem{EvCa89} K.E. Evans, B. D. Caddock, 
{\it Microporous materials with negative PoissonÕs ratio: II. Mechanisms and interpretation} J. Phys. D., Appl. Phys. {\bf 22}, 1883 (1989).
\bibitem{Al85} R.F. Almgren, 
{\it An isotropic three dimensional structure with PoissonÕs ratio 1Ú4 2 1}, J. Elasticity {\bf 15}, 427 (1985).
\bibitem{Ev89} K.E. Evans, 
{\it Tensile network microstructures exhibiting negative PoissonÕs ratio} J. Phys. D., Appl. Phys. {\bf 22}, 1870 (1989).
\bibitem{Bl05} R. Blumenfeld, 
{\it Auxetic strains Ð insight from iso-auxetic materials}, Molecular Simulations {\bf 31}, 867 (2005).
\bibitem{Cosserat} E. Cosserat, F. Cosserat. 
{\it Theorie des Corps Deformables} (A. Hermann et Fils, 1909).
\bibitem{Wietal96} J.P. Wittmer, P. Claudin, M.E. Cates and J.-P. Bouchaud, 
Nature {\bf 382}, 336 (1996).
\bibitem{Wietal97} J.P. Wittmer, M.E. Cates and P. Claudin, 
J. Phys. I (France), {\bf 7}, 39 (1997).
\bibitem{Caetal98} M. E. Cates, J. P. Wittmer, J.-P. Bouchaud and P. Claudin, 
Phys. Rev. Lett. {\bf 81}, 1841 (1998).
\bibitem{BaBl02} R. C. Ball and R. Blumenfeld, 
{\it The stress field in granular systems: Loop forces and potential formulation}, Phys. Rev. Lett. {\bf 88} 115505 (2002).
\bibitem{Bl04a} R. Blumenfeld, 
{\it Stresses in granular systems and emergence of force chains}, Phys. Rev. Lett. {\bf 93}, 108301 (2004) 
\bibitem{Bl05} R. Blumenfeld, 
{\it Stresses in two-dimensional isostatic granular systems: Exact solutions}, New J. Phys. {\bf 9}, (2007) 160.
\bibitem{Gretal05} J.N. Grima, A. Alderson, K.E. Evans. 
{\it Auxetic behaviour from rotating rigid units}, Phys. Stat. Sol. (b) {\bf 242}, 561 (2005).
\bibitem{Euler} This relation was discovered by Euler circa 1750. See P. R. Cromwell, 
 {\it Polyhedra}, pp. 189-190 (Cambridge University Press, Cambridge 1997).
\bibitem{BlEd03} R. Blumenfeld and Sam F. Edwards, 
 {\it Granular entropy: Explicit calculations for planar assemblies} Phys. Rev. Lett. {\it 90}, 114303 (2003).
\bibitem{BlEd06} R. Blumenfeld and Sam F. Edwards, 
 {\it Geometric partition functions of cellular systems: Explicit calculation of the entropy in two and three dimensions}Eur. Phys. J. {\bf E 19}, 23 (2006).
\bibitem{Blaus07} R. Blumenfeld, in {\it Lecture Notes in Complex Systems Vol. 8: Granular and Complex Materials}, pp 43, eds. T. Aste, A. Tordesillas and T. D. Matteo (World Scientific 2007, Singapore).
\bibitem{BaBl03}  R. C. Ball and R. Blumenfeld, 
 {\it From plasticity to renormalisation group}, Phil. Trans. R. Soc. Lond. {\bf 360}, 731 (2003).
 \bibitem{Bl04b}  R. Blumenfeld, {\it Stress in planar cellular solids: Coarse-graining the constitutive equation}, Physica {\bf A 336}, 361 (2004).
\bibitem{Venant} C. L. M. H.  Navier, {\it R\'{e}sum\'{e} des Lexcons sur lÕapplication de la Me\'{e}chanique, 3\'{e}me edition avec des notes et des appendices par A. J. C. Barr\'{e} de Saint-Venant. Dunod}, (Paris 1864).


\end{thebibliography}
\end{document}